\documentclass[12pt]{article}
\usepackage{latexsym,amssymb,amsmath}
\begin{document}

\title
{Self-dual Yang--Mills fields \\
in pseudoeuclidean spaces}
\author
{E.K. Loginov\footnote{Research supported by RFBR Grant 04-02-16324;
E-mail: loginov@ivanovo.ac.ru}\\
Physics Department, Ivanovo State University\\
Ermaka St. 39, Ivanovo, 153025, Russia}
\date{}
\maketitle

\begin{abstract}
The self-duality Yang--Mills equations in pseudoeuclidean spaces
of dimensions $d\leq 8$ are investigated. New classes of solutions
of the equations are found. Extended solutions to the $D=10$,
$N=1$ supergravity and super Yang-Mills equations are constructed
from these solutions.
\end{abstract}

\section{Introduction}

In 1983 Corrigan et al.~[1] have proposed a generalization of the
self-dual Yang--Mills equations in dimension $d>4$:
\begin{equation}
f_{mnps}F^{ps}=\lambda F_{mn},
\end{equation}
where the numerical tensor $f_{mnps}$ is completely antisymmetric
and $\lambda=const$ is a non-zero eigenvalue. By the Bianchi
identity $D_{[p}F_{mn]}=0$, it follows that any solution of (1) is a 
solution of the Yang--Mills equations $[D^m,F_{mn}]=0$. Some of these solutions have found in~[2].
\par
The many-dimensional Yang--Mills equations appear in the
low-energy effective theory of superstring and
supermembrane~[3,4]. In addition, there is a hope that Higgs
fields and super\-symmetry can be understood through dimensional
reduction from $d>4$ dimensions down to $d=4$~[5].
\par
The paper is organized as follows. Section 2 contains well-known
facts about Cayley-Dickson algebras and connected with them Lie
algebras. In Sections 3 and 4 the self-duality Yang--Mills equations in
pseudoeuclidean spaces of dimensions $d\leq 8$ are investigated.
In Section 5 extended solutions to the $D=10$, $N=1$ supergravity
and super Yang-Mills equations are constructed from these
solutions.

\section{Cayley-Dickson algebras}

Let us recall that the algebra $A$ satisfying the identities
\begin{equation}
x^{2}y=x(xy),\qquad yx^{2}=(yx)x
\end{equation}
is called alternative. It is obvious that any associative algebra
is alternative. The most important example of nonassociative
alternative algebra is Cayley-Dickson algebra. Let us recall its
construction~(see~[6]).
\par
Let $A$ be an algebra with an involution $x\to\bar x$ over a field
$F$ of characteristic $\ne 2$. Given a nonzero $\alpha\in F$ we
define a multiplication on the vector space $(A,\alpha)=A\oplus A$
by
\begin{equation}
(x_1,y_1)(x_2,y_2)=(x_1x_2-\alpha \bar y_2y_1,y_2x_1+y_1\bar x_2).
\notag
\end{equation}
This makes $(A,\alpha)$ an algebra over $F$. It is clear that $A$
is isomorphically embedded into $(A,\alpha)$ and
$dim(A,\alpha)=2dimA$. Let $e=(0,1)$. Then $e^2=-\alpha$ end
$(A,\alpha)=A\oplus Ae$. Given any $z=x+ye$ in $(A,\alpha)$ we
suppose $\bar z=\bar x-ye$. Then the mapping $z\to\bar z$ is an
involution in $(A,\alpha)$.
\par
Starting with the base field $F$ the Cayley-Dickson construction
leads to the following sequence of alternative algebras:
\par
1) $F$, the base field.
\par
2) ${\mathbb C}(\alpha)=(F,\alpha)$, a field if $x^2+\alpha$ is
the irreducible polynomial over $F$; otherwise, ${\mathbb
C}(\alpha)\simeq F\oplus F$.
\par
3) ${\mathbb H}(\alpha,\beta)=({\mathbb C}(\alpha),\beta)$, a
generalized quaternion algebra. This algebra is associative but
not commutative.
\par
4) ${\mathbb O}(\alpha,\beta,\gamma)=({\mathbb
H}(\alpha,\beta),\gamma)$, a Cayley-Dickson algebra. Since this
algebra is not associative the Cayley-Dickson construction ends
here.
\par
The algebras in 1) -- 4) are called composition. Any of them has
the nondegenerate quadratic form (norm) $n(x)=x\bar x$, such that
$n(xy)=n(x)n(y)$. In particular, over the field ${\mathbb R}$ of
real numbers, the above construction gives 3 split algebras (e.g.,
if $\alpha=\beta=\gamma=-1$) and 4 division algebras (if
$\alpha=\beta=\gamma=1$): the fields of real ${\mathbb R}$ and
complex ${\mathbb C}$ numbers, the algebras of quaternions
${\mathbb H}$ and octonions ${\mathbb O}$, taken with the
Euclidean norm $n(x)$. Note also that any simple nonassociative
alternative algebra is isomorphic to Cayley-Dickson algebra
${\mathbb O}(\alpha,\beta,\gamma)$.
\smallskip\par
Let $A$ be Cayley-Dickson algebra and $x\in A$. Denote by $R_x$
and $L_x$ the operators of right and left multiplication in $A$
$$
R_x:a\to ax, \qquad L_x:a\to xa.
$$
 It follows from (2) that
\begin{equation}
R_{ab}-R_aR_b=[R_a,L_b]=[L_a,R_b]=L_{ba}-L_aL_b.
\end{equation}
Consider the Lie algebra ${\cal L}(A)$ generated by all operators
$R_x$ and $L_x$ in $A$. Choose in ${\cal L}(A)$ the subspaces
$R(A)$, $S(A)$, and $D(A)$ generated by the operators $R_{x}$,
$S_{x}=R_x+2L_x$, and $2D_{x,y}=[S_{x},S_{y}]+S_{[x,y]}$
respectively. Using (3), it is easy to prove that
\begin{align}
3[R_{x},R_{y}]&=D_{x,y}+S_{[x,y]},\\
[R_{x},S_y]&=R_{[x,y]},\\
[R_{x},D_{y,z}]&=R_{[x,y,z]},\\
[S_{x},S_{y}]&=D_{x,y}-S_{[x,y]},\\
[S_{x},D_{y,z}]&=S_{[x,y,z]},\\
[D_{x,y},D_{z,t}]&=D_{[x,z,t],y}+D_{x,[y,z,t]},
\end{align}
where $[x,y,z]=[x,[y,z]]-[y,[z,x]]-[z,[x,y]]$. It follows from
(4)--(9) that the algebra ${\cal L}(A)$ is decomposed in the
direct sum
$$
{\cal L}(A)=R(A)\oplus S(A)\oplus D(A)
$$
of the Lie subalgebras $D(A)$, $D(A)\oplus S(A)$ and the vector
space $R(A)$.
\par
In particular, if $A$ is a real division algebra, then $D(A)$ and
$D(A)\oplus S(A)$ are isomorphic to the compact Lie algebras $g_2$
and $so(7)$ respectively. If $A$ is a real split algebra, then
$D(A)$ and $D(A)\oplus S(A)$ are isomorphic to noncompact Lie
algebras $g'_2$ and $so(4,3)$.

\section{Self-dual solutions in $d=8$}

Let  $A$ be a real linear space equipped with a nondenerate
symmetric metric $g$ of signature $(8,0)$ or $(4,4)$. Choose the
basis \{$1,e_{1},...,e_{7}\}$ in $A$ such that
\begin{equation}
g=\text{diag}(1,-\alpha,-\beta,\alpha\beta,-\gamma,\alpha\gamma,\beta\gamma,-\alpha\beta\gamma),
\end{equation}
where $\alpha,\beta,\gamma=\pm1$. Define the multiplication
\begin{equation}
e_{i}e_{j}=-g_{ij}+c_{ij}{}^{k}e_{k},
\end{equation}
where the structural constants $c_{ijk}=g_{ks}c_{ij}{}^{s}$ are
completely antisymmetric and different from 0 only if
\begin{equation}
c_{123}=c_{145}=c_{167}=c_{246}=c_{275}=c_{374}=c_{365}=1.
\end{equation}
The multiplication (11) transform $A$ into a linear algebra. It
can easily be checked that the algebra $A\simeq\mathbb
O(\alpha,\beta,\gamma)$. In the basic \{$1,e_{1},...,e_{7}$\} the
operators
\begin{equation}
R_{e_i}=e_{i0}+\frac12c_{i}{}^{jk}e_{jk},\qquad
L_{e_i}=e_{i0}-\frac12c_{i}{}^{jk}e_{jk},
\end{equation}
where $e_{ij}$ are generators of the Lie algebra ${\cal L}(A)$
satisfying the switching relations
\begin{equation}
[e_{mn},e_{ps}]=g_{mp}e_{ns}-g_{ms}e_{np}-g_{np}e_{ms}+g_{ns}e_{mp}.
\end{equation}
\par
Now, let $G$ be a matrix Lie group constructed by the Lie algebra
$D(A)\oplus S(A)$. In the space $A$ equipped with the metric (10)
we define the completely antisymmetric $G$-invariant tensor
$f_{mnps}$~(cp.~[7]):
\begin{align}
f_{ijk0}&=c_{ijk},
\notag\\
f_{ijkl}&=g_{il}g_{jk}-g_{ik}g_{jl}+c_{ijm}c_{kl}{}^{m}, \notag
\end{align}
where $i,j,k,l\ne 0$. Representing the nonzero components of
$f_{mnps}$ in the explicit form
\begin{align}
f_{0123}=f_{0145}=f_{0167}=f_{0246}=f_{0275}=f_{0374}=f_{0365}=1,
\notag\\
f_{4567}=f_{2367}=f_{2345}=f_{1357}=f_{1364}=f_{1265}=f_{1274}=1,
\notag
\end{align}
we see that the tensor $f_{mnps}$ satisfies the identity
\begin{equation}
f_{mnij}f_{ps}{}^{ij}=6(g_{mp}g_{ns}-g_{ms}g_{np})+4f_{mnps}.
\end{equation}
Define the projectors $\tilde f_{mnps}$ of ${\cal L}(A)$ onto the
subalgebra $D(A)\oplus S(A)$ and its generators $\tilde f_{mn}$ by
\begin{align}
\tilde
f_{mnps}&=\frac38(g_{mp}g_{ns}-g_{ms}g_{np})-\frac18f_{mnps},
\notag\\
\tilde f_{mn}&=\tilde f_{mn}{}^{ij}e_{ij}. \notag
\end{align}
It follows from (15) that
\begin{align}
f_{mnij}\tilde f_{ps}{}^{ij}&=-2\tilde f_{mnps},\\
f_{mnij}\tilde f^{ij}&=-2\tilde f_{mn}.
\end{align}
Using  the identities (7)--(9) and (13), we get the switching
relations
\begin{equation}
[\tilde f_{mn},\tilde f_{ps}]=\frac34(\tilde f_{m[p}g_{s]n}-\tilde
f_{n[p}g_{s]m})-\frac18(f_{mn}{}^{k}{}_{[p}\tilde
f_{s]k}-f_{ps}{}^{k}{}_{[m}\tilde f_{n]k}).
\end{equation}
\par
Now we can find solutions of (1). We choose the ansatz~(cp.~[2])
\begin{equation}
A_m(x)=\frac43\frac{\tilde f_{mi}x^{i}}{\rho^2+r^{2}},
\end{equation}
where $r^{2}=x_{k}x^{k}$ and $\rho\in{\mathbb R}$. Using the switching
relations (18), we get
\begin{equation}
F_{mn}(x)=-\frac49\frac{(6\rho^2+3r^2)\tilde f_{mn}+8\tilde
f_{mni}{}^s\tilde f_{sj}x^ix^j}{(\rho^2+r^2)^2}.
\end{equation}
It follows from (16)--(17) that the tensor $F_{mn}$ is self-dual.
If the metric (10) is Euclidean, then we have the well-known
solution of equations (1)~(see.~[2]). If the metric (10) is
pseudoeuclidean, then we have a new solution.

\section{Solutions in $d<8$}

Now we'll find solutions of the self-duality equations in
dimension $d<8$. If $B_{\alpha}$ is a subalgebra of the real
Cayley-Dickson algebra $A$, then the subgroup $H_{\alpha}$ of
automorphisms of $A$ leaving fixed the element of $B_{\alpha}$ is
called the Galois group of $A$ over $B_{\alpha}$. Description of
these groups is known~[8]. In particular, if $A$ is the real
division algebra and $B_{1}\simeq\mathbb R$, $B_{2}\simeq\mathbb
C$, $B_{3}\simeq\mathbb H$, then
$$
G\simeq Spin(7),\quad H_{1}\simeq G_{2},\quad H_{2}\simeq
SU(3),\quad H_{3}\simeq SU(2).
$$
If $A$ is the real split algebra, then for the same choice of
$B_{i}$,
$$
G\simeq Spin(4,3),\quad H_{1}\simeq G'_{2},\quad H_{2}\simeq
SU(2,1),\quad H_{3}\simeq SU(1,1).
$$
Obviously, the orthogonal complement $B_{\alpha}^{\perp}$ of
$B_{\alpha}$ in $A$ is  $H_{\alpha}$-invariant subspace of
dimension $d_{\alpha}=8-dimH_{\alpha}$. Now it is easy to
construct a completely antisymmetric $H_{\alpha}$-invariant
$d_{\alpha}$-tensor $f^{\alpha}_{mnps}$. It is a projection of the
$d$-tensor $f_{mnps}\in \Lambda^{4}(A)$ onto the subspace
$\Lambda^{4} (B_{\alpha})$. We can choose nonzero components of
$f_{mnps}$ in the form
\begin{align}
f^{1}_{4567}=f^{1}_{2367}=f^{1}_{1274}=f^{1}_{1357}
=f^{1}_{1364}=f^{1}_{1265}=f^{1}_{2345}&=1,
\notag\\
f^{2}_{1364}=f^{2}_{1265}=f^{2}_{2345}&=1,
\notag\\
f^{3}_{2345}&=1. \notag
\end{align}
Now we can easily get analogues of the identities (15)--(18). In
particular, the switching relations (18) takes the form
\begin{equation}
[\tilde f^{\alpha}_{mn},\tilde f^{\alpha}_{ps}]
=\frac{3-\alpha}{4-\alpha}(\tilde f^{\alpha}_{m[p}g_{s]n} -\tilde
f^{\alpha}_{n[p}g_{s]m})
-\frac1{8-2\alpha}(f^{\alpha}_{mn}{}^{k}{}_{[p}\tilde
f^{\alpha}_{s]k}-f^{\alpha}_{ps}{}^{k}{}_{[m}\tilde
f^{\alpha}_{n]k}). \notag
\end{equation}
Note that if we choose the ansatz $A_{m}(x)$ in the form
\begin{equation}
A_m(x)=k\frac{\tilde f^{\alpha}_{mi}x^{i}}{\rho^2+r^{2}},
\notag
\end{equation}
then the corresponding field strength $F_{mn}$ is not self-dual for $\alpha=2$. On
the contrary, if $\alpha=1$ or $\alpha=3$, then choice of
$A_{m}(x)$ in the form (21) reduce to a self-dual field strength.
For example, if $\alpha=1$, then $k=3/2$ and
\begin{equation}
F_{mn}(x)=-\frac32\frac{(2\rho^2+r^2)\tilde f^{1}_{mn}
+3\tilde f^{1}_{mni}{}^s\tilde
f^{1}_{sj}x^ix^j}{(\rho^2+r^2)^2}.
\end{equation}
For Euclidean metric this solution is known~(see.~[2]). For
pseudoeuclidean metric we have a new solution. For $\alpha=3$ we
have the well-known instanton solution~[9] or its noncompact
analogue.
\par
Note that in $d=4$ there exist another solution of the Yang--Mills equations. It depends on coordinates of the Minkowski space. Indeed, we choose the ansatz $A_{m}(x)$ in the form
\begin{equation}
A_m(x)=\frac{2e_{mn}x^{n}}{\lambda^2+x_{k}x^{k}},
\end{equation}
where $e_{mn}$ are generators of the Lie algebra $so(p,q)$ satisfying the  relations $so(p,q)$. Then the field strength
\begin{equation}
F_{mn}(x)=\frac{-4\lambda^2e_{mn}}{(\lambda^2+x_{k}x^{k})^{2}},
\end{equation}
and
\begin{equation}
\partial^{m}F_{mn}+[A^{m},F_{mn}]
=\frac{8\lambda^{2}e_{mn}x^{m}}
{(\lambda^{2}+x_{k}x^{k})^{3}}(4-\delta_{i}^{i}).
\notag
\end{equation}
Hence the anzatz (22) satisfies the Yang--Mills equations if $p+q=4$. If $\mid p-q\mid=4$ or $0$, then the algebra $so(p,q)$ is isomorphic to the direct sum $so(3)\oplus so(3)$ or $so(2,1)\oplus so(2,1)$ of proper subalgebras. Therefore projecting $A_m(x)$ on these subalgebras, we again get the instanton solution~[9] or its noncompact analogue. If $\mid p-q\mid=2$, then the algebra $so(p,q)$ is simple. In this case the solution (23) of the Yang--Mills equations is not self-dual. 

\section{Extended supersymmetric solutions}

Let now show that the above instanton solutions can be extended to
solutions of the $D=10$, $N=1$ supergravity and super Yang-Mills
equations. Consider the purely bosonic sector of the theory
\begin{equation}
S=\frac{1}{2k^2}\int d^{10}x\sqrt{-g}e^{2\phi}\left(R+4(\nabla)^2
-\frac{1}{3}H^2-\frac{\alpha'}{30}\text{Tr}(F^2)\right).
\end{equation}
Rather than directly solve the equations of motion for this
action, it is much more convenient to look for bosonic backgrounds
which are annihilated by some of $N=1$ supersymmetry
transformations. This requires that in ten dimensions there exist
at least one Majorana-Weyl spinor $\epsilon$ such that the super
symmetry variations
\begin{align}
\delta\chi&=F_{MN}\Gamma^{MN}\epsilon,\notag\\  
\delta\lambda&=\left(\Gamma^{M}\partial_{M}\phi-\frac16H_{MNP}\Gamma^{MNP}\right)\epsilon,\notag\\
\delta\psi_{M}&=\left(\partial_{M}+\frac14\left(\omega_{M}^{AB}-H_{M}^{AB}\right)\Gamma_{AB}\right)\epsilon\notag
\end{align}
of the fermionic fields vanish for such solutions. We will
construct a simple ansatz for the bosonic fields which does just
this~(cp.~[3]).
\par
First, we choose $\epsilon$ to be $Spin(4,3)$-singlet of Majorana-Weyl 
spinor of $SO(5,5)$. Denote world indices of the eight-dimensional 
subspace indices by $\mu,\nu=1,\dots,8$ and the corresponding tangent 
space indices by $m,n=1,\dots,8$. We assume that no fields depend on the
coordinates with indices $M,N=0,9$. It follows that
$$
\tilde f_{mnps}\Gamma^{ps}\epsilon=\tilde f_{mn}\epsilon=0.
$$
Using the expression (20) for tensor field strength $F_{mn}$, we
see that the supersymmetry variation $\delta\chi$ vanishes. 
\par
Further, we choose the antisymmetric tensor field strength $H_{mnp}$ and
metric tensor $g$ in the form
\begin{align}
H_{mnp}&=\frac17f_{mnps}\partial^s\phi,\\
g_{\mu\nu}&=e^{(6/7)\phi}g_{0\mu\nu},\notag
\end{align}
where $g_0$ is the pseudoeuclidean metric (10). 
Using the identities
$$
f_{mnps}\Gamma^{mnp}=42\Gamma_{s}
$$
and the explicit form of spin connectedness
$$
\omega_{\mu mn}=\frac37(\delta_{\mu m}\partial_{n}\phi-\delta_{\mu
n}\partial_{m}\phi),
$$
we prove that the variations $\delta\lambda$ and  $\delta\psi_M$
also vanish for any $\phi(x)$.
\par
It follows from the Bianchi identity
$$
\partial_{[m}H_{nps]}=-\alpha'\text{Tr}_{8}F_{[mn}F_{ps]}
$$
that the tensor field strength
\begin{equation}
H_{mnp}=-\alpha'\frac{3\rho^{2}
+r^{2}}{9(\rho^{2}+r^{2})^{3}}f_{mnps}x^{s}.
\end{equation}
If we compare (26) with (25), we find
\begin{equation}
e^{(6/7)\phi}=e^{(6/7)\phi_0}
+\alpha'\frac{2\rho^{2}+r^{2}}{3(\rho^{2}+r^{2})^2},
\end{equation}
where $\phi_{0}$ is the value of the dilaton $\phi$ on $\infty$.
Similarly, if we choose $G'_2$-singlet of Majorana-Weyl spinor of
$SO(5,5)$ and use the expression (21) for the tensor field
strength $F_{mn}$, we get an analog of the solution (27).

\end{document}